\documentclass{ws-ijmpa}
\begin{document}

\markboth{R.~S.~Decca {\it et al.}}{Capacitance Measurements and
Electrostatic Calibrations}

%
\catchline{}{}{}{}{}
%

\title{CAPACITANCE MEASUREMENTS AND ELECTROSTATIC
{\protect \\}CALIBRATIONS IN EXPERIMENTS MEASURING
{\protect \\}THE CASIMIR FORCE}

\author{R.~S.~DECCA,${}^1$
E.~FISCHBACH,${}^2$ G.~L.~KLIMCHITSKAYA,${}^3$ D.~E.~KRAUSE,${}^{4,2}$
D.~L\'{O}PEZ,${}^5${\protect \\}  U.~MOHIDEEN${}^6$
and V.~M.~MOSTEPANENKO${}^7$}

\address{${}^1$Department of Physics, Indiana University-Purdue
University Indianapolis, Indianapolis, Indiana 46202, USA
{\protect \\}
${}^2$Department of Physics, Purdue University, West Lafayette, Indiana
47907, USA
{\protect \\}
${}^3$North-West Technical University,
 Millionnaya Street 5, St.Petersburg,
191065, Russia
{\protect \\}
${}^4$Physics Department, Wabash College, Crawfordsville, Indiana 47933, USA
{\protect \\}
${}^5$Center for Nanoscale Materials, Argonne National Laboratory,
Argonne, Illinois 60439, USA
{\protect \\}
${}^6$Department of Physics and Astronomy, University of California,
Riverside, CA 92521, USA
{\protect \\}
${}^7$Noncommercial Partnership
``Scientific Instruments'',
Tverskaya Street 11, Moscow,
103905, Russia
}

\maketitle

\begin{history}
\received{2 June 2011}
\end{history}

\begin{abstract}
We discuss the possibility of determining the properties and quality
of spherical surfaces used in precise experiments with the help
of capacitance measurements.
The results of this kind measurements for the lens-plane and
sphere-plane, Au coated surfaces are compared
with theoretical predictions from various
models of perfect and broken sphericity.
It is shown that capacitance measurements
are incapable of discriminating between models of perfect and
modified centimeter-size
spherical surfaces in an experiment demonstrating the
anomalous scaling law for the electric force. Claims to the contrary
in the recent literature are explained by the use of improper
comparison. The data from capacitance measurements in an experiment
measuring the Casimir force by means of a micromechanical torsional
oscillator employing micrometer-size spheres
are shown to be in excellent agreement with theoretical
predictions using the model of a perfect spherical surface.

\end{abstract}

\ccode{PACS numbers: 68.35.Ct, 68.47.De, 84.37.+q}

\section{Introduction}

Starting from 1997, measurements of the Casimir force\cite{1}
between surfaces of a gold-coated spherical lens or
sphere and a plate has attracted
widespread attention in areas ranging from nanotechnology to
constraining predictions of fundamental physical theories
beyond the Standard Model (see  a overview\cite{2}
of the subject).
The Casimir force arises between surfaces of two electrically neutral
closely spaced bodies. It is the specific case of the van der
Waals force\cite{3} when the width of a gap between the
surfaces is sufficiently large so that the retardation
effects become important. Both the van der Waals and the
Casimir force are quantum phenomena caused by the existence
of electromagnetic fluctuations.\cite{4} They are closely
related to other fluctuation phenomena such as, for instance,
radiative heat transfer.\cite{5,6}

Measuring small forces, such as the Casimir force,
is a complicated scientific and
technological problem.
A detailed analysis of all performed experiments on
measuring the Casimir force can be found in
review.\cite{7}
An important constituent of any Casimir force measurement is the
electrostatic calibration. This allows an independent determination
of some vitally important parameters (the absolute separation
between two closely spa\-ced bodies, the residual potential
difference between their surfaces, spring constant etc.) by
fitting the electric force arising between
a lens or a sphere and a plate
under an applied voltage to the known force law.
The experimental procedures for electrostatic calibrations were
discussed in detail in the pioneering papers.\cite{8}\cdash\cite{10}
A basic assumption used was that the electric force is given
by the exact formula of electrostatics obtained for an ideal
metal sphere above an ideal metal plane.\cite{11}
It is important to bear
in mind\cite{7}
that the typical sphere radii used in these experiments were about
$R=100\,\mu$m and separation distances between the sphere and
the plate were of order $d=100\,$nm.

Kim et al.\cite{12} reconsidered electrostatic calibration in
the sphere-plane geometry by replacing the small sphere with a
spherical lens of centimeter-size radius of curvature $R$ at a very
close separation from the plate, $d\geq 30\,$nm.
Both test bodies were covered with gold layers.
The experimental data obtained from the electrostatic calibration
demonstrated an anomalous dependence of the gradient of the
electric force on separation $\sim R/d^{1.7}$ instead
of $R/d^2$, as given by the main contribution to the
exact formula. This result was discussed in
Refs.~\refcite{13,14}. Specifically, it was
demonstrated\cite{13} that the data for the electrostatic force
between a sphere and a plate in the separation range
$d\geq 100\,$nm followed the standard electrostatic law.
It was concluded\cite{13} that the observation\cite{12}
is not universal.  A model of a
modified geometry of a spherical surface was proposed\cite{14} which
provides possible explanation for the anomalous behavior
of the electric force.\cite{12}
However, according to Ref.~\refcite{15},
the model, which explains the anomalous behavior of the
gradient of electric force, ``is hard to reconcile with
the measurement of the capacitance versus distance that better
follows the behavior expected for a sphere with a single
radius of curvature''.

In this paper we discuss the possibility of determining properties and quality
of spherical surfaces using
capacitance measurements in the
electrostatic calibration of the Casimir setup.
In Sec.~2, a brief summary of the main results for the
capacitance in the sphere-plane geometry is presented.
In Sec.~3, the capacitance measurements\cite{15,16}
 are compared with theoretical
predictions following from the model of a perfect spherical
surface,\cite{15,16} and from the
alternating model with a modified
surface\cite{14} providing the explanation of the anomalous
behavior of the electric force.\cite{12}
We show that the conclusion,\cite{15} that the
capacitance measurements are compatible with the model of
perfect sphere, but incompatible with the model of
modified spherical surface, is based on an improper comparison.
Sec.~4 is devoted to the capacitance measurements
in the experiment on the dynamic determination of the
Casimir pressure by means of a micromechanical torsional
oscillator. We present the experimental data for the
capacitance in the configuration of a  Au-coated
micrometer-size
sapphire sphere above an Au-coated polysilicon plate, and
demonstrate excellent agreement with theoretical
results of electrostatics for a perfectly shaped sphere.
 Sec.~5 contains our
conclusions and discussion.

\section{Capacitance in a Sphere-Plate Geometry}

We consider an ideal metal sphere of radius $R$ at a
separation $d$ above an ideal metal plane. The exact
expression for the electrical capacitance in such a
configuration is given\cite{17} by
\begin{equation}
C(d)=4\pi\epsilon_0 R\,{\rm sinh}\,\alpha\sum_{n=1}^{\infty}
\frac{1}{{\rm sinh}(n\alpha)},
\label{eq1}
\end{equation}
\noindent
where ${\rm cosh}\,\alpha=1+d/R$ and $\epsilon_0$ is the
permittivity of a vacuum. When a voltage $V$ is applied to
the plane while the sphere is kept grounded, the electrostatic
energy of sphere-plane interaction is
\begin{equation}
E(d)=-\frac{1}{2}C(d)(V-V_0)^2,
\label{eq2}
\end{equation}
\noindent
where $V_0$ is the residual potential difference when both
bodies are grounded. Using Eq.~(\ref{eq1}) one obtains the
exact expression for the electric force acting between a sphere
and a plane\cite{11}
\begin{equation}
F(d)=-\frac{\partial E}{\partial d}=\frac{(V-V_0)^2}{2}
\frac{\partial C(d)}{\partial d}
=
2\pi\epsilon_0 (V-V_0)^2
\sum_{n=1}^{\infty}
\frac{{\rm coth}\,\alpha-n{\rm coth}(n\alpha)}{{\rm sinh}(n\alpha)}.
\label{eq3}
\end{equation}
\noindent
In the limiting case of small separations, $d/R\ll 1$, an
approximate expression for the capacitance (\ref{eq1}) was
obtained\cite{18}
\begin{equation}
C(d)\approx 2\pi\epsilon_0 R\left(\ln\frac{R}{d}+\ln\,2+
\frac{23}{20}+\frac{\tau}{63}\right),
\label{eq4}
\end{equation}
\noindent
where $\tau$ is a number such that $0\leq\tau\leq 1$.

 A very precise expansion for the electric force
(\ref{eq3}) within a wide region of parameters was obtained\cite{19}
in the form of the following expansion:
\begin{equation}
F(d)=-2\pi\epsilon_0(V-V_0)^2\sum_{k=-1}^{6}c_k
\left(\frac{d}{R}\right)^k,
\label{eq5}
\end{equation}
\noindent
where
$c_{-1}=0.5,\quad c_0=-1.18260$, $c_1=22.2375$,
$c_2=-571.366$, $c_3=9592.45$, $c_4=-90200.5$,
$c_5=383084$, $c_6=-300357$.
The first term of (\ref{eq5}) with $k=-1$ coincides with the
force calculated using the proximity force
approximation\cite{20} (PFA).
The accuracy of (\ref{eq5}) depends on the separation region
and on the value of a sphere radius. For example, at
$0.5\,\mu\mbox{m}\leq d \leq 4\,\mu$m for $R\approx 150\,\mu$m
(the parameters of the capacitance measurements in the
experiment using a micromechanical torsional
oscillator\cite{21,22}
discussed in Sec.~3) the computational results obtained
using (\ref{eq5}) coincide with those obtained from
(\ref{eq3}) to within 0.06\%. With larger $R$, the agreement
between (\ref{eq3}) and (\ref{eq5}) becomes better.

Using the second equality in (\ref{eq3}) and (\ref{eq5}),
we obtain
\begin{equation}
\frac{\partial C(d)}{\partial d}=\frac{2F(d)}{(V-V_0)^2}=
-4\pi\epsilon_0\sum_{k=-1}^{6}c_k
\left(\frac{d}{R}\right)^k.
\label{eq7}
\end{equation}
\noindent
The integration of (\ref{eq7}) leads to
\begin{equation}
C(d)=4\pi\epsilon_0 R\left[c_{-1}\ln\frac{R}{d}+\tilde{c}-
\sum_{k=0}^{6}\frac{c_k}{k+1}\left(\frac{d}{R}\right)^{k+1}
\right].
\label{eq8}
\end{equation}
\noindent
Here, the integration constant $\tilde{c}$ can be found from the
comparison with Eq.~(\ref{eq4})
\begin{equation}
\tilde{c}=\frac{1}{2}\ln\,2+\frac{23}{40}+\frac{\tau}{126}.
\label{eq9}
\end{equation}
\noindent

We emphasize that the coefficient of the leading, logarithmic, term
in (\ref{eq4}) and (\ref{eq8}) is fixed theoretically up to the
error in the measurement of the sphere radius. During electric
measurements, there are wires connected to the sphere,
the plate, and neighboring parts of the setup leading to
parasitic capacitances. These are discussed in Sec.~4.

\section{Capacitance Measurements and the Anomaly in Electrostatic
Calibrations}

The anomalous behavior of the electric force was observed\cite{12}
in the configuration of an Au-coated spherical lens of radius
of curvature
$R=30.9\pm 0.15\,$mm spaced more than $d=30\,$nm above an Au-coated
Si plate. As was mentioned in Sec.~1, the
experimental data\cite{12}
 demonstrated that in the separation region from
30 to 100\,nm the gradient of the electric force varies with
separation as $\sim R/d^{1.7}$ instead of the expected behavior
$\sim R/d^2$.
Such an
anomaly might be explained\cite{14} by the local modification of the lens
surface due to the presence of two sectors with curvature radii
$R_1=1.6R=49.4\,$mm, $R_2=30\,\mu$m, and heights $H=250\,$nm
and $h=8\,$nm, respectively. Such local modifications of the lens
surface are easily allowed by the specifications provided by the
manufacturer.
Using the PFA, this modified geometry of the
sphere leads to the following modified electric force\cite{14}
\begin{equation}
F^{\rm mod}(d)=-\pi\epsilon_0(V-V_0)^2\!\left(\frac{R_2}{d}+
\frac{R_1-R_2}{d+h}-\frac{R_1-R}{d+h+H}\right)\!.
\label{eq10}
\end{equation}
\noindent
This expression is in very good agreement with the anomalous
behavior of the electric force\cite{12}
within the entire measurement range from 30 to 100\,nm
(see Fig.~2 of  Ref.~\refcite{14}). The corresponding
capacitance for a sphere with
the modified geometry above the plate is obtained from
Eq.~(\ref{eq10})
by integration with respect to $d$
\begin{equation}
C^{\rm mod}(d)=2\pi\epsilon_0\left[R_2\ln\frac{R_2}{d}+
(R_1-R_2)\ln\frac{R_1-R_2}{d+h}
-(R_1-R)\ln\frac{R_1-R}{d+h+H}\right]
+\tilde{C},
\label{eq11}
\end{equation}
\noindent
where $\tilde{C}$ is the integration constant.

The approximate values of $C^{\rm mod}$ in (\ref{eq11}) at small
separations in the region from 30 to 100\,nm can be calculated using
the formula
\begin{equation}
C^{\rm mod}(d)\approx A_1^{\rm mod}+A_3^{\rm mod}
\left(\frac{d}{R}\right)^{0.3}
+\tilde{C},
\label{eq12}
\end{equation}
\noindent
where $A_1^{\rm mod}=32.804\,$pF, $A_3^{\rm mod}=-360.48\,$pF.
At large separations above $1\,\mu$m, the asymptotic behavior of
$C^{\rm mod}$ is given by
\begin{equation}
C^{\rm mod}(d)\approx 2\pi\epsilon_0 R\ln\frac{R}{d}
 +
2\pi\epsilon_0 R\frac{H}{d}\left(\frac{R_1-R}{R}-\frac{R-R_2}{R}\,
\frac{h}{H}\right)
+\tilde{C}.
\label{eq13}
\end{equation}
\noindent
In Fig.~1(a), the solid line shows the exact dependence of $C^{\rm mod}$
on $d$ in accordance to (\ref{eq11}),
the approximate dependence at short separations in  (\ref{eq12})
is shown by the dotted line, and the   large
separation behavior (\ref{eq13}) is indicated by the dashed line
(all functions are plotted with $\tilde{C}=0$).
\begin{figure*}[t]
\vspace*{-2.3cm}
\centerline{\hspace*{1.1cm}\psfig{file=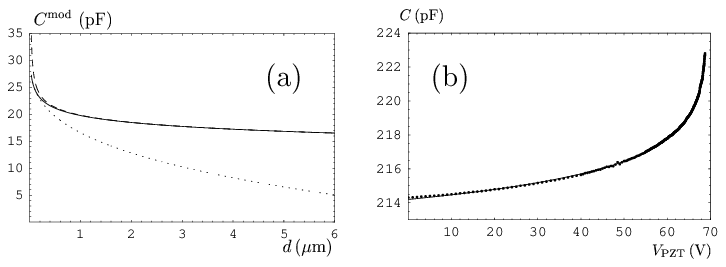,width=21cm}}
\vspace*{-22.7cm}
\caption{
(a) The capacitance (\ref{eq11})
of a spherical lens with a mo\-di\-fied geometry
above a plane (the solid line) and its approximate expressions
(\ref{eq13}) at large
and (\ref{eq12}) at small separations (the dashed and dotted
lines, respectively)
versus separation.
(b) The measurement data for the capacitance in the lens-plane
geometry versus voltage applied to the piezo are indicated as dots.
The solid line is the fit to the data using the model of the modified
geometry of the lens surface.
}
\end{figure*}

According to Refs.~\refcite{15,16}, the expression (\ref{eq11})
for $C^{\rm mod}$ ``is hard to reconcile with the measurement of the
capacitance versus distance that better follows the behavior expected
for a surface with a single radius of curvature''. This claim is in
contradiction with electrostatics because the electric force is connected
with the derivative of a capacitance in accordance with the second
equality in
(\ref{eq3}). Keeping this in mind, it seems improbable that the data
from the electric force measurements demonstrate anomalous deviations
from the force scaling law given by the main contribution to
Eqs.~(\ref{eq3}) and (\ref{eq5}), while the data of the capacitance
measurements in the same experiment were in agreement with (\ref{eq3}).

To resolve this puzzle, we repeated the comparison
of the experimental data for the capacitance
measurements\cite{15} with the models of both
modified and perfect spherical
surfaces. We begin with the case
of a modified spherical surface. First, we note that
the data of capacitance measurements were compared\cite{15}
not with our exact
equation (\ref{eq11}) but with the following approximate expression for it
 \begin{equation}
\tilde{C}^{\rm mod}(d)= \tilde{A}_1^{\rm mod}+\tilde{A}_3^{\rm mod}
d^{0.3}.
\label{eq14}
\end{equation}
\noindent
Another approximate expression containing an additional term of the form
$\tilde{A}_2d$ was also suggested.\cite{15} We do not consider it
here because the graphical information in Fig.~1 of Ref.~\refcite{15} is
related to (\ref{eq14}).

It is seen that (\ref{eq14}) is of the same form as the approximate
expression (\ref{eq12})
which is valid at short separations only. However,
 the function (\ref{eq14}) was fitted\cite{15,16} to the
experimental data for the capacitance measurements over a wide separation
region up to $6\,\mu$m. By doing so
Refs.~\refcite{15,16} have used the
following  relationship between the separation
$d$ and the voltage $V_{\rm PZT}$ applied to the piezo:
$
d=\beta(V_{\rm PZT}^{0}-V_{\rm PZT})$,
$\beta=87\pm 2\,\mbox{nm/V}$.
The fit was performed with the use of $n=363$ data points
$\>(V_{{\rm PZT},i},C_i)$ over the range of $V_{\rm PZT}$ from 0 to 68.76\,V.
These data points are shown as dots in Fig.~1(b). The values of three
fitting parameters are $V_{\rm PZT}^0=68.43\pm 0.05\,$V,
 $\tilde{A}_1^{\rm mod}=222.96\pm 0.04\,$pF, and
 $\tilde{A}_3^{\rm mod}=-(346.2\pm 1)\,$pF/m${}^{0.3}$
(the dimension pF/m indicated\cite{15} is presumably a misprint).

Note that the resulting values of the fitting parameters are physically
unacceptable. First, for the 17 largest experimental voltages
$V_{{\rm PZT},i}$ (of the total number of 363) the related separations
 turn out to be negative. Next, the value
of the coefficient $\tilde{A}_3^{\rm mod}$
obtained from the fit
 leads to
$\tilde{A}_3^{\rm mod}R^{0.3}=-121.99\,$pF, i.e., differs by a factor
of about 3 from the value ${A}_3^{\rm mod}$ in the approximate
expression (\ref{eq12}) applicable at small separations. One can
conclude that the function (\ref{eq14}) (shown as the dashed line in
Fig.~1 of Ref.~\refcite{15}) is in poor agreement with the
data of capacitance measurements. Hence,
the conclusion\cite{15} made on this basis,
that the capacitance (\ref{eq11})
for the model of a modified spherical surface is in worse agreement
with the data than the model of a perfect sphere,
is not supported. It follows that
 the exact expression in (\ref{eq11}) cannot be
approximated by (\ref{eq14}) over a wide separation region.

To determine the extent of agreement between the exact
expression (\ref{eq11}) and
the experimental data we have performed a direct fit with two fitting
parameters $\tilde{C}$ and $V_{\rm PZT}^{0}$.
This fit results in  $\tilde{C}=197.69\pm 0.01\,$pF and
$V_{\rm PZT}^{0}=69.93\pm 0.02\,$V (see below for
a description of the fitting
procedure used). In Fig.~1(b) we plot the corresponding $C^{\rm mod}$
versus $V_{\rm PZT}$ as the solid line. As is seen in Fig.~1(b),
the experimental data are in much better agreement with (\ref{eq11})
than with (\ref{eq14}) (compare with the dashed line in
Fig.~1 of Ref.~\refcite{15}). For example, the values of the capacitance
at $V_{\rm PZT}=0$ computed from (\ref{eq14}) and (\ref{eq11}) are
equal to 213.59 and 214.20\,pF, respectively. This should be compared
with the experimental value of $C=214.313\pm 0.0015\,$pF.

We next  compare the results of our fit using (\ref{eq11})
with the results of the fit using the model of an ideal
spherical surface.\cite{15,16}. For the theoretical dependence of the
capacitance on separation, Refs.~\refcite{15,16} use
\begin{equation}
C^{\rm id}(d)=A_1^{\rm id}+A_3^{\rm id}\ln\frac{R}{d}.
\label{eq16}
\end{equation}
\noindent
Note that, in disagreement with (\ref{eq4}), the negative value
$A_3^{\rm id}=-2\pi\epsilon_0R$ is quoted. In addition,
the numerical value of $A_3^{\rm id}$ is indicated as
$A_3^{\rm id}=-(1.72\pm 0.02)\,$pF. However, direct substitution
of $\epsilon_0$ and $R$ given in the beginning of this section leads
to $A_3^{\rm id}=-(1.719\pm 0.008)\,$pF, i.e., the error of
$A_3^{\rm id}$ is overestimated.

\begin{figure*}[t]
\vspace*{-5.8cm}
\centerline{\hspace*{-2.1cm}\psfig{file=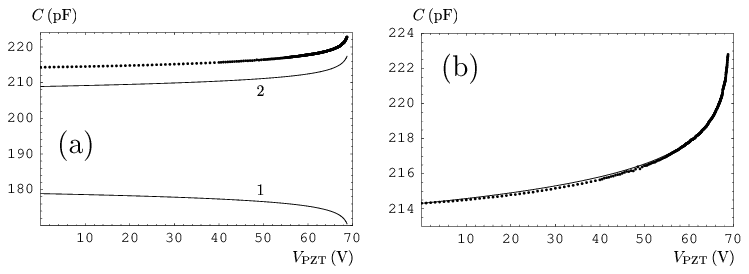,width=21cm}}
\vspace*{-18.8cm}
\caption{
Data for the capacitance measurements in the lens-plane
geometry versus voltage applied to the piezo are indicated as dots.
(a) Lines 1 and 2 and (b) the solid line are the fits to the data
 using the model of an ideal spherical surface with various
parameters (see text for further discussion).
}
\end{figure*}
In spite of the fact that $A_3^{\rm id}$ is known theoretically, the fit
of Eq.~(\ref{eq16}) was performed with three fitting parameters,
$A_1^{\rm id}$, $A_3^{\rm id}$ and $V_{\rm PZT}^{0}$, with the
result\cite{15} $A_1^{\rm id}=193.9\pm 0.2\,$pF,
$A_3^{\rm id}=-(1.757\pm 0.002)\,$pF, and $V_{\rm PZT}^{0}=69.31\pm 0.02\,$V.
In Fig.~2(a) we present the capacitance $C^{\rm id}$ given
by  (\ref{eq16}) as a function of
$V_{\rm PZT}$ (solid line 1). In the same figure the experimental data
are shown as dots. It is seen that the results of the fit\cite{15}
 are incorrect. By choosing the
positive sign of $A_3^{\rm id}$ according to (\ref{eq4}),
we arrive at the solid line 2 in Fig.~2(a) which also
disagrees significantly with
the data. However, the computational results\cite{15,16}
can be reproduced if one replaces the value of
the fitting parameter\cite{15,16}
$A_1^{\rm id}=193.9\pm 0.2\,$pF  with
$A_1^{\rm id}=199.3\pm 0.2\,$pF. With this replacement
we plot the resulting solid line in Fig.~2(b) where the experimental
data are once again shown as dots. Figures 1(b) and 2(b)
qualitatively demonstrate the extent of agreement between the data of
capacitance measurements and theoretical predictions from the model
of a modified and an ideal spherical surface, respectively.

The fit of Eq.~(\ref{eq11}) comparing the case of a modified spherical
surface to the experimental data was performed using the maximum
likelihood method, i.e., by the minimization of the function\cite{23}
\begin{equation}
M=\sum_{i=1}^{n}\frac{[C_i-C^{\rm mod}(d_i)]^2}{\sigma_{C_i}^2}.
\label{eq17}
\end{equation}
\noindent
We recall that $n=363$ and $C^{\rm mod}(d_i)$ are given in (\ref{eq11}).
The measured values of the capacitances at the applied voltages
$V_{{\rm PZT},i}$ and their experimental errors are denoted as $C_i$ and
$\sigma_{C_i}$, respectively. The values of $\sigma_{C_i}$
 are presented in Fig.~3(a). As was mentioned
in Sec.~3, this is a fit with $r=2$ parameters, $\tilde{C}$ and
$V_{\rm PZT}^{0}$. The values of these parameters providing the minimum
value of $M$, which is usually referred to as $\chi^2$, were listed above.
\begin{figure*}[t]
\vspace*{-2.3cm}
\centerline{\hspace*{1.1cm}\psfig{file=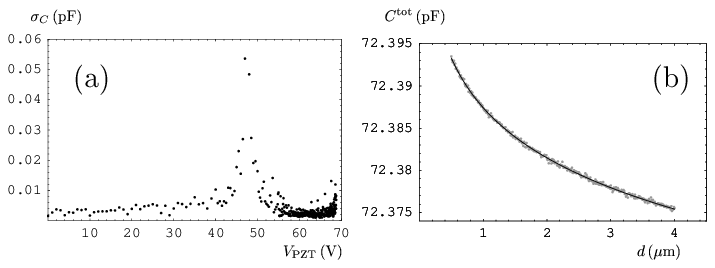,width=21cm}}
\vspace*{-22.3cm}
\caption{
(a) The measurement errors of the capacitance in Ref.~15
versus voltage applied to the piezo are indicated as dots.
(b) The measurement data for the total capacitance in a
sphere-plane geometry for the experiment using a micromechanical
torsional oscillator are indicated as gray dots. The solid line is
the fit to the data using the model of a perfect spherical surface.
}
\end{figure*}

The measure of agreement between the experimental data and some fitting
function is the so-called reduced-$\chi^2$ equal to $\chi^2/f$, where
$f=n-r$ is the number of degrees of freedom (in our case $f=361$).
Using these definitions one can calculate that for the model of
a modified spherical surface [see Eq.~(\ref{eq11}) and Fig.~1(b)]
the reduced $\chi^2$ is approximately equal to 715.
For the model of
an ideal sphere\cite{15} [see Eq.~(\ref{eq16}) and
Fig.~2(b)] (\ref{eq17}) leads to a reduced $\chi^2$ equal to 1100.
So large values of the reduced $\chi^2$ are explained by
 the smallness of $\sigma_{C_i}$ in Fig.~3(a).
We emphasize that significantly different values for the reduced
$\chi^2$ are given in Ref.~\refcite{15}. Thus, for the model of an ideal
sphere a reduced $\chi^2$ equal to 2.9 is reported.\cite{15}
For the modified spherical surface described  by Eq.~(\ref{eq14}),
instead of the exact function
(\ref{eq11}), a reduced $\chi^2$ equal to 77.4
is given.\cite{15} Both these values are not
reproducible.

The large values of the reduced $\chi^2$ obtained by us
mean that the fit is in fact
not satisfactory. It is well
known\cite{23} that if the resulting value of $\chi^2$ is much larger
than $f$, one should carefully check all assumptions on which the choice
of the fitting function is based.
For all fits mentioned above, the probability of obtaining
 a larger value of $\chi^2$ is negligibly
small. This means that the data in fact are not related to the model
being used in the fit.
Thus, the failure of the model
with an ideal geometrical shape is not surprising  because spherical
surfaces of centimeter-size radii inevitably deviate from perfect
sphericity.\cite{14} Our model (\ref{eq11}) takes into account
deviations from sphericity only in the close vicinity of
the  lens bottom point.
This is sufficient to correctly describe the anomalous behavior\cite{12}
of the electric force at short separations between a lens and a plate
below 100\,nm, but not sufficient to fit the measurement data for the
capacitance over much wider separation regions. The latter are influenced
by the deviations of the lens surface from sphericity over a much larger
area. In fact
both the force and the capacitance are very sensitive to the variation
of geometry,
and hence attempts to fit the experimental data without
a careful study of the
surface topography are not productive. In the next section we
analyze the results of the capacitance measurements for the most
precise experiment on the Casimir force\cite{21,22} using a small sphere
of the best achievable quality.

\section{Capacitance Measurements in the Experiment Using a
Micromechanical Torsional Oscillator}

The experiment\cite{21,22} on the dynamic determination of the Casimir
pressure using a micrometer-size
 sphere oscillating in the vertical direction
near a plate suspended at two opposite points by serpentine springs
is the most precise experiment in Casimir physics.
This is the only measurement of the Casimir interaction where random
errors are much smaller than systematic errors.
Previous publications\cite{21,22} contain
the results of electrostatic calibrations using the electric
force and the indirect measurements of the Casimir pressure
(the electrostatic calibration is considered in more
detail in Ref.~\refcite{24}).
Here, we report the results of the capacitance measurements in the same
experiment and their comparison with theory.

Capacitance measurements were performed
using the same setup as was used to measure the Casimir force.
Data reported here were acquired at the same time
as  the data for
the electrostatic force and the Casimir force were measured.\cite{21,22}
The main part of the setup was an Au-coated sapphire sphere
above an Au-coated polysilicon microelectromechanical torsional
oscillator (MTO). The sphere had a radius $R = 151.3 \pm 0.2\,\mu$m
and the plate had dimensions of $500\times 500\,\mu\mbox{m}^2$. The sphere was
glued to an Au-covered optical fiber.
The purpose of the fiber is to directly measure the
changes in separation between the end of the fiber and the
platform that holds the MTO (see, e.g., Refs.~\refcite{7,10,24}
for the schematic of the setup).

Capacitance measurements were performed with a
{AH2700A} Capacitance Bridge
 at 10 kHz in the bridge ba\-lance
mode (i.e. the capacitance to be measured was ba\-lanced against calibrated
capacitances in the bridge to provide a nearly-null output). The AC voltage
between the sphere and the MTO was maintained at 30 mV, and each measurement
was performed over a 0.1 Hz bandwidth.
The MTO, as well as the surrounding metallic structures in
the system, were kept at the same potential, while the potential of
the sphere-fiber assembly was varied sinusoidally at a frequency
$f \simeq 10\,$kHz, much larger than the resonant frequency of the
MTO. Under these conditions, the oscillator is stationary, and the
capacitance can be obtained directly. The total capacitance of the
system $C^{\rm tot}$ is found as
$C^{\rm tot}(d)=C(d) + C^{\rm p}(d)$, where $C^{\rm p}(d)$ is
the parasitic capacitance, determined mainly by the capacitance
between wires, and the capacitance between the fiber and the
platform (the capacitance between the sphere-fiber assembly and
the rest of the system is negligible). While the capacitance
between wires is independent of separation, the
capacitance between the end of the fiber and the platform is a
function of their separation.
Altogether, 351 measurements of the capacitance $C_i^{\rm tot}$
have been performed within the separation region from 500.5\,nm
to 4000.2\,nm. Absolute separations $d_i$ were measured\cite{10,21,22}
with an absolute error $\Delta d=0.6\,$nm.
 Within the separation region
of capacitance measurements $\Delta d\ll d_i$ holds. Because of
this, the error in the measurement of absolute separations does not
play any role and can be neglected in the fit performed below.
The error of capacitance measurements in this experiment
$\sigma_i\equiv\sigma=2\times 10^{-4}\,$pF does not depend on
separation.

The results of capacitance measurements are shown in Fig.~3(b) as gray dots.
These results were fitted to the total capacitance of the system
given by the sum of the exact expression
(\ref{eq1}) in a sphere-plate configuration,
and the parasitic capacitance. The latter is given by
$C^{\rm p}(d)=A^{\rm w}+A^{\rm pl}/z_{\rm meas}$ where $z_{\rm meas}$
is the separation between the end of the optical fiber and the platform.
It is measured using two-color interferometer.
Here, the quantity $A^{\rm w}$ is separation independent, and is
dominated by the capacitance between wires. The term
$A^{\rm pl}/z_{\rm meas}$ models the capacitance of the plane capacitor
formed by the optical fiber and the platform.
This term can be represented as $A_1^{\rm pl}-A_2^{\rm pl}d/D$.
For this purpose we use the expression
$z_{\rm meas}=d+D+b\theta$, where
$\theta$ is a negligibly small angle of rotation of the plate,
$b$ is the lever arm, and $D$ is the sum of the distance between the
bottom of the optical fiber and the bottom of the sphere and the distance
between the platform and the top of the plate (see Fig.~1 in
Ref.~\refcite{10}). Note that the quantity $D$ does not depend on the
separation $d$ between the sphere and the plate\cite{2,7,21,24} and
$d\ll D$.
Introducing the notations $\tilde{A}_1=A^{\rm w}+A_1^{\rm pl}$
and $\tilde{A}_2=A_2^{\rm pl}/D$, we can write the parasitic capacitance in
the form
\begin{equation}
C^{\rm p}(d)=\tilde{A}_1- \tilde{A}_2d.
\label{eq18}
\end{equation}
\noindent
The fit of the results of capacitance measurements to the total
theoretical capacitance $C^{\rm tot}(d)$, which is equal to the sum
of (\ref{eq1}) and (\ref{eq18}), was performed by the
minimization of the function (\ref{eq17}) with two unknown parameters
$\tilde{A}_1$ and $\tilde{A}_2$. The resulting values of these
parameters providing a minimum value to the quantity (\ref{eq17}) are
$\tilde{A}_1=72.32971\pm 0.00002\,$pF and
$\tilde{A}_2=(2.18\pm 0.10)\times 10^{-4}\,\mbox{pF/$\mu$m}$.
The plot of
theoretical $C^{\rm tot}(d)$ versus separation is shown
in Fig.~3(b) by the solid line. Taking into account that $n=351$ and
the fit has $r=2$ parameters, we obtain for the number of degrees of
freedom $f=349$. The resulting reduced $\chi^2$ is equal to 0.7
which leads to a probability to obtain not a smaller value
of $\chi^2$ very close to unity.
We emphasize that the fit was performed with respect to only two
parameters of the parasitic capacitance which is determined by
uncontrolled random factors. Thus, not obtaining  a smaller
value of $\chi^2$ in each next repetition of the measurement is
really highly probable.
The excellent agreement of our capacitance data with the model
of a perfect spherical surface is seen in Fig.~3(b) where most of
deviations of the data dots from the theoretical curve are in
the limits of experimental errors.
The seemingly larger scatter of the experimental
dots around the solid line in Fig.~3(b), in comparison with Figs.~1(b) and
2(b), is explained by the different ranges of the variation of
capacitance. In Figs.~1(b) and 2(b) capacitance varies by about 10\,pF over
the separation region of a few micrometers, but in Fig.~3(b) the variation
of capacitance over the similar separation region is only 0.02\,pF.

In the above fit we have used the exact expression
(\ref{eq1}) for the capacitance
in the configuration of a sphere above a plane. To discuss the
usefulness of capacitance measurements, it is interesting to compare
our results with those obtained when some approximate
expression is used in the fit. For example,
if we use the PFA, the capacitance is given by the leading, logarithmic,
contribution on the right-hand side of (\ref{eq4}) and the
respective electric force by the first term (with $k=-1$) on the
right-hand side of (\ref{eq5}). The parasitic capacitance is
represented by (\ref{eq18}) as before. When we now perform the fitting
procedure of the same data but using the simplified function
\begin{equation}
C^{\rm PFA}(d)=2\pi\epsilon_0R\ln\frac{R}{d}+\tilde{A}_1- \tilde{A}_2d,
\label{eq19}
\end{equation}
\noindent
the resulting values of the coefficients are
$\tilde{A}_1=72.34530\pm 0.00002\,$pF and
$\tilde{A}_2=(1.11\pm 0.10)\times 10^{-4}\,\mbox{pF/$\mu$m}$.
They are slightly different from the case when the exact expression for
the capacitance in the sphere-plane configuration has been used.
What is important, however, is that the reduced $\chi^2$ for
Eq.~(\ref{eq19}) is equal to 0.7, i.e., is the same as in the case of the
exact equation (\ref{eq1}). This means that it may be not possible to
uniquely choose a preferable model when fitting several theoretical
expressions
to the experimental data of capacitance measurements
with fitting parameters owing to the parasitic capacitances.

\begin{table}[b]
\tbl{The values of the capacitances
and electric forces [the latter are normalized on the factor
$-(V-V_0)^2$] in the sphere-plane geometry ($R=151.3\,\mu$m)
computed by using the exact formulas (columns 2 and 5),
the PFA (columns 3 and 6), and by the perturbative expansions
(columns 4 and 7) at different separations (column 1).}
{\begin{tabular}{@{}ccccccc@{}}\toprule
&\multicolumn{3}{c}{$C\,$(fF)}&
\multicolumn{3}{c}{$-F/(V-V_0)^2\,$(nN/V${}^2$)} \\
\cline{2-4}\cline{5-7}
$d\,(\mu\mbox{m})$
& exact &PFA & expansion&exact &PFA & expansion  \\
\colrule
0.5 & 63.71 & 48.08 & 63.60 & 8.35023 & 8.41721 & 8.35518 \\
1.0 & 57.94 & 42.25 & 57.70 & 4.14806 & 4.20860 & 4.14975 \\
1.5 & 54.58 & 38.84 & 54.23 & 2.74897 & 2.80574 & 2.74956 \\
2.0 & 52.22 & 36.41 & 51.76 & 2.05022 & 2.10430 & 2.05040 \\
2.5 & 50.39 & 34.54 & 49.83 & 1.63144 & 1.68344 & 1.63147 \\
3.0 & 48.91 & 33.00 & 48.24 & 1.35256 & 1.40287 & 1.35256 \\
3.5 & 47.66 & 31.70 & 46.89 & 1.15359 & 1.20246 & 1.15358 \\
4.0 & 46.59 & 30.58 & 45.72 & 1.00454 & 1.05215 & 1.00453 \\
\botrule
\end{tabular}\label{tab1}}
\end{table}
This does not mean, however, that the more exact and
the less exact approximate analytical
expressions for the capacitance and the electric force are equally
applicable in electrostatic calibrations. For comparison purposes,
in Table~1 we present the computational results for the capacitances
(column 2--4) and the electric forces normalized by the factor
$-(V-V_0)^2$ (columns 5--7) at different separations (column 1).
The values of the capacitances and electric forces are calculated using
the exact expressions
(\ref{eq1}) and (\ref{eq3}) (columns 2 and 5, respectively),
by the PFA, i.e., by the first terms in (\ref{eq4}) and (\ref{eq5})
(columns 3 and 6), and using the more precise expansions in the powers of
$d/R$ in (\ref{eq8}) and (\ref{eq5}) (columns 4 and 7).
As is seen in Table~1 (columns 2 and 3), for the capacitance, the PFA
reproduces the exact results with a rather large relative error.
When the separation varies from 0.5 to $4\,\mu$m, this error increases
from 24.5\% to 34.4\%. Such large errors are explained by the fact that
the PFA does not take into account the constant term in the capacitance
(\ref{eq4}). When such a constant $\tilde{A}_1$ is added
to model the parasitic capacitances in (\ref{eq19}), and is
determined from the fit, it compensates for the missing
separation-independent contribution in the theoretical expression
provided by the PFA. As to the perturbative expansion for the capacitance
in column 4, it is in much better agreement with the exact results in
column 2. The respective relative error increases from 0.17\% to only
1.86\% when the separation increases from 0.5 to $4\,\mu$m.

With respect to the electric force measurements, the PFA (column 6)
is in much better agreement with the exact results presented in column 5.
Thus, with the increase of $d$ from 0.5 to $4\,\mu$m, the relative errors of
the force values computed using PFA increase from 0.8\% to only
4.7\%. Such a good agreement is explained by the fact that the constant
contribution to the capacitance, omitted in the PFA, does not influence
the electric force. Even better agreement is found in Table~1 when
comparing the force values computed using the perturbative expansion
(column 7) with the exact values (column 5). Here, the relative error
varies from 0.06\% to 0.001\% when the separation increases
from 0.5 to $4\,\mu$m. Once again, it can be seen that the perturbative
expansion for the electric force is much more exact than for the
capacitance. This permits us to state that the primary role in the
electrostatic calibrations of Casimir setups should be given to
the electric force.

\section{Conclusions and Discussion}

In the foregoing we have considered
the possibility of characterizing the properties of spherical
surfaces using
 capacitance measurements as a part
of the electrostatic calibrations in experiments on measuring the Casimir
force. This subject was stimulated by
the anomalous scaling law  reported\cite{12} for the electric force
in the configuration of a centimeter-size
lens above a plane plate, and the
subsequent discussion.\cite{13}\cdash\cite{16} Keeping in mind the importance
of electrostatic calibrations for the determination of
the precision of the
experimental results used in numerous applications of the Casimir force,
a conclusive explanation of the puzzle is highly desirable.
An attempt at such an explanation was undertaken\cite{14}
by suggesting the model of a modified lens surface which reproduces the
anomalous scaling law for the force.\cite{12}
This explanation was disputed\cite{15,16} by claiming that
the suggested model is in much worse agreement with the results of
capacitance measurements than the model of perfect spherical lens
surface.
Here, we demonstrate that the objections\cite{15,16} against
the proposed model\cite{14} are invalid. As shown in Sec.~3,
instead of comparing the exact expression for the capacitance\cite{14}
with the corresponding experimental data,
the comparison of the data with another function
was made\cite{15} which does
not reproduce the behavior of the exact expression.\cite{14}
We have shown that the exact capacitance\cite{14} is in better
agreement with the data than the one for an ideal
spherical surface.\cite{15,16}
At the same time, we have also shown that
the agreement with the data of both models of a lens surface modified
near the closest point to the plate and of an ideal spherical surface
is not satisfactory. This can be explained by the role of more
irregularities distributed over a larger area of the surface
of a centimeter size sphere
which inevitably contribute to the capacitance.
Recently it was recognized\cite{25} that local geometrical
deformations of the surface can really lead to an anomalous
electrostatic force and this should be taken into account in
future experiments. Furthermore, it was shown\cite{26} that
bubbles and pits which are invariably present on lens surfaces
of centimeter-size radii of curvature result in large
uncontrollable corrections to the Casimir force. This makes
fundamentally flawed all measurements of the Casimir force
employing centimeter-size spherical lenses (see,
for instance, Ref.~\refcite{27}).

To add  important new information to this discussion, in Sec.~4 we
have presented new experimental data on the capacitance measurements in
the Casimir setup using a micromechanical torsional
oscillator.\cite{10,21,22,24}
This setup includes a perfectly shaped sapphire sphere
of a radius 200 times smaller than the lens radius in
Refs.~\refcite{12,15,16}.
The experimental data were carefully compared with the exact
expression for the capacitance in a sphere-plane configuration.
Different approximate representations for it, as discussed in Sec.~2,
were also analyzed. It was shown that the  data for the
capacitance measurement in the setup using a micromechanical torsional
oscillator are in excellent agreement with the model of a perfect
spherical surface. This provides additional confirmation for the high
quality of the Au-coated micrometer-size
sapphire sphere used in that experiment.

One additional conclusion obtained in Sec.~4
is that by using the capacitance
measurements and by fitting them to different theoretical expressions
it may be difficult to conclude which expression is in better
agreement with data due to the existence of parasitic capacitances.
The reason is that a contribution from the parasitic capacitances
is unavoidably contained in any capacitance measurement
both in macro and microscales.
This contribution
cannot be calculated theoretically with sufficient accuracy
and its parameters are determined from the fit.
It can be concluded that in spite of the fact that capacitance and
force are connected by Eq.~(\ref{eq3})
the capacitance measurements in
the Casimir setups should be considered as only an auxiliary tool
providing an opportunity to confirm the good quality
of a micrometer-size spherical
surface used.

\section*{Acknowledgments}
The authors are grateful to R.\ Onofrio for providing data for
$C$ and $\sigma_{C}$ versus $V_{\rm PZT}$ in his experiment.
R.S.D.~acknowledges NSF support through Grant
No.~PHY-0701236, and LANL support through Contract No.~49423--001--07.
D.L.\ and R.S.D.~acknowledge support from DARPA Crant
No.~09--Y557.
E.F. was supported in part by the DOE under Grant No.~DE-76ER071428.
G.L.K., V.M.M.\ and U.M. were supported by the NSF Grant
No.\ PHY0970161
(computation of capacitances) and DOE Grant
No.\ DEF010204ER46131 (statistical analysis).


\end{document}